\documentclass[english]{article}
\usepackage[latin9]{inputenc}
\usepackage{geometry}
\usepackage{amsmath}
\usepackage{amssymb}
\usepackage{setspace}
\usepackage{bbold} 
\newcommand{\UM}{\mathbb{1}}
\newcommand{\R}{\mathbb{R}}
\newcommand{\C}{\mathbb{C}}
\newcommand{\Hbb}{\mathbb{H}}
\newcommand{\Obb}{\mathbb{O}}
\makeatletter
\newcommand{\lyxaddress}[1]{
\par {\raggedright #1
\vspace{1.4em}
\noindent\par}
}

\makeatother

\usepackage{babel}
\begin{document}

\title{Octonionic formulation of the fully symmetric Maxwell's equations in $3+1$ dimensions}

\author{K.\ Pushpa and J.\ C.\ A.\ Barata}

\maketitle

\lyxaddress{\begin{center}
Instituto de F\'{\i}sica \\
Universidade de S\~ao Paulo\\
Caixa Postal 66 318 \\
 05315 970 São Paulo. SP. Brasil 
\par\end{center}}

\lyxaddress{\begin{center}
Email: 
pushpa@if.usp.br and 
pushpakalauni60@yahoo.co.in\;,\\
jbarata@if.usp.br\\

\par\end{center}}
\begin{abstract}
  Using octonions, more specifically, using a $4\times4$ matrix
  representation of octonions obtained with the help of algebraic
  properties of quaternions, we obtain the fully symmetric Maxwell's
  equations (Maxwell's equations with electric and magnetic charges and
  currents which is invariant under duality transformations). 
\end{abstract}

\section{Introduction}

The algebra of octonions $\Obb$ forms the largest normed division
algebra over the real numbers $\R$, complex numbers $\C$, and
quaternions $\Hbb$ (see e.g. \cite{key-1}). The algebra of octonions
$\Obb$ is an algebraic structure defined on the
8\textendash{}dimensional real vector space $\R^8$ and is a
non-associative extension of the algebra of quaternions $\Hbb$. These
composition algebras are responsible for many important mathematical
structures of interest for physicists \cite{key-2,key-3}.  Octonions
have been used in various ways to describe properties of quantum
mechanics \cite{key-4}, field theory \cite{key-5,key-6}, gauge theory
\cite{key-7,key-8}. Electromagnetism has been expressed on octonionic
algebras \cite{key-9,key-10,key-11,key-12}, split octonion algebra
\cite{key-13} and octonion Dirac equations \cite{key-14,key-15}.

In the presence of sources, the usual Maxwell's equations are neither
symmetric nor invariant with respect to the duality transformation
between electric and magnetic fields. Dirac \cite{key-16} proposed the
existence of magnetic monopoles to symmetrise the Maxwell's equations.
He also argued that the existence of an isolated magnetic charge would
imply the quantisation of electric charge. Dirac \cite{key-17} also
described the applications of quaternions to Lorentz transformations.
Symmetric Maxwell's equations are invariant under Lorentz and duality
transformation. t\textquoteright{}Hooft \cite{key-18} and Polyakov
\cite{key-19} gave the idea that the magnetic monopoles may be found
in Yang-Mill\textquoteright{}s theory. In the present study, we
describe the fully symmetric Maxwell's equations by using the algebraic
properties of quaternions and octonions. This study will give an
octonionic representation of fully symmetric Maxwell's equations as a
single equation.

\section{Octonions}

The octonionic algebra $\Obb$ is an eight dimensional algebra over the
field of the real numbers $\R$. Its elements can be represented as
\begin{align}
O\; = \; & e_{0}y_{0}+e_{1}y_{1}+e_{2}y_{2}+e_{3}y_{3}+e_{4}y_{4}+e_{5}y_{5}+e_{6}y_{6}+e_{7}y_{7}\nonumber \\
\; = \; & e_{0}y_{0}+\sum_{a=1}^{7}e_{a}y_{a}\; ,
\label{eq:1}
\end{align}
where $y_{k}$ $(k=0,\, 1,\, \ldots\, ,7)$ are real numbers, $e_{a}$
$(a=1,\, 2,\, \ldots ,\, 7)$ are imaginary octonion units and $e_{0}$ is the
multiplicative unit element. The set of octets $(e_{0},\, e_{1},\,
e_{2},\, e_{3},\, e_{4},\, e_{5},\, e_{6},\, e_{7})$ is known as the
octonion basis and its elements satisfy the following multiplication
rules:
\begin{eqnarray}
e_{0} & = & 1 \; , \quad \mbox{{\it i.e.,}}\quad e_{0}e_{a}\;=\; e_{a}e_{0}\;=\;e_{a}\nonumber \\
e_{a}e_{b} & = & -\delta_{ab}e_{0}+f_{abc}e_{c}, \; (a,\, b,\, c\, =\, 1,\, 2,\, \ldots , \, 7)\;,\label{eq:2}
\end{eqnarray}
where $\delta_{ab}$ is the usual Kr\"{o}necker symbol and the
structure constants $f_{abc}$ are completely antisymmetric, take
the value $1$ for the following combinations:
\begin{gather}
f_{abc}=+1 \;\; \forall(abc)\; =\; (123),\; (471),\; (257),\; (165),\; (624),\; (543),\; (736)\; ,\label{eq:3}
\end{gather}
and vanish otherwise.
The octonion basis elements satisfy the following additional relations:
\begin{eqnarray}
\left[e_{a},\,\, e_{b}\right] & = & 2\, f_{abc}\, e_{c}\; ,\nonumber \\
\left\{ e_{a},\,\, e_{b}\right\}  & = & -2\,\delta_{ab}e_{0}\;,\nonumber \\
e_{a}(\, e_{b}\, e_{c}) & \neq & (e_{a}\, e_{b}\,)\, e_{c}\;,\label{eq:4}
\end{eqnarray}
where the brackets $[\;,\;]$ and $\{\;,\;\}$ denote, as usual,
the commutator and the anti-commutator, respectively. 

Notice that the sub-algebra generated by $(e_{0},\, e_{7})$ is isomorphic
to the algebra $\C$ of the complex numbers while the sub-algebra generated
by $(e_{0},\, e_{1},\, e_{2},\, e_{3})$ is isomorphic to the algebra
$\Hbb$ of quaternions.

\section{Representation of octonions in terms of 4$\times$4 matrix}

The algebra of octonions is homeomorphic to a four-dimensional algebra
over the complex numbers. In order to see this, consider that the algebra
of the complex numbers is isomorphic to the sub-algebra of $\Obb$ generated
by the elements ($e_{0},\, e_{7})$, since $(e_{7})^{2}=-e_{0}$. Now,
any octonion $O=e_{0}y_{0}+e_{1}y_{1}+e_{2}y_{2}+e_{3}y_{3}+e_{4}y_{4}+e_{5}y_{5}+e_{6}y_{6}+e_{7}y_{7}\in \Obb$
can be written as
\begin{align}
O & =\left(y_{0}+e_{7}y_{7}\right)e_{0}+\left(y_{1}+e_{7}y_{4}\right)e_{1}+\left(y_{2}+e_{7}y_{5}\right)e_{2}+\left(y_{3}+e_{7}y_{6}\right)e_{3}
\;.
\label{eq:5}
\end{align}
Denoting
\begin{eqnarray}
Y_{0} & = & y_{0}+e_{7}y_{7}\; ,\nonumber \\
Y_{j} & = & y_{j}+e_{7}y_{j+3}\; ,\quad \forall j=1,\,2,\,3\;,\label{eq:6}
\end{eqnarray}
equation (\ref{eq:5}) becomes
\begin{align}
O & =Y_{0}e_{0}+Y_{1}e_{1}+Y_{2}e_{2}+Y_{3}e_{3} \; .\label{eq:7}
\end{align}
Equation (\ref{eq:7}) represents octonions in terms of elements
of the sub-algebra of quaternions (the sub-algebra generated by $(e_{0},\, e_{1},\, e_{2},\, e_{3})$,
with \textquotedblleft{}coefficients\textquotedblright{} in the sub-algebra
$(e_{0},\, e_{7})$, isomorphic to the algebra of complex numbers.
It is well known that the algebra of quaternions is isomorphic to
the algebra of Pauli matrices through the identification 
\begin{align}
e_{0}\;=\;\UM \; \quad \mbox{ and }\quad & e_{j}\;= \;-i\sigma_{j}\;, \quad \forall j\,=\,1,\,2,\,3,\label{eq:8}
\end{align}
where the unit matrix $\UM$ and  the Pauli matrices $\sigma_{j}$ are given, as usual, by
\begin{gather}
\UM=\left[\begin{array}{cc}
1 & 0\\
0 & 1
\end{array}\right]\; ,
\quad \sigma_{1}=\left[\begin{array}{cc}
0 & 1\\
1 & 0
\end{array}\right]\; ,
\quad \sigma_{2}=\left[\begin{array}{cc}
0 & -i\\
i & 0
\end{array}\right]\; ,
\quad \sigma_{3}=\left[\begin{array}{cc}
1 & 0\\
0 & -1
\end{array}\right]\;.
\label{eq:9}
\end{gather}
In order to obtain a representation of $\Hbb$ in terms of real matrices
one has to proceed in the following way. Define $J=\left[\begin{array}{cc}
0 & 1\\
-1 & 0
\end{array}\right]$ .
Since $J^2=-\UM$, we may replace the imaginary number $i$ by $J$ in
(\ref{eq:8}), through the following identifications:
$$
e_0 \; = \; \UM \otimes \UM, \;\; 
e_1 \; = \; -J \otimes \sigma_1, \;\; 
e_2 \; = \; \UM \otimes (-i\sigma_2) \; = \; -\UM\otimes J , \;\; 
\mbox{ and }
e_2 \; = \;  -J \otimes \sigma_3.
$$
This provides following identification of the elements of the
quaternionic basis with \underline{real} $4\times 4$ matrices:
$$
e_0 \; = \; 
\begin{pmatrix}
\UM & 0 \\ 0 & \UM
\end{pmatrix}
, \;\;
e_1 \; = \; 
\begin{pmatrix}
0 & -J \\ -J & 0
\end{pmatrix}
, \;\;
e_2 \; = \; 
\begin{pmatrix}
0 & -\UM \\ \UM & 0
\end{pmatrix}
, \;\;
e_3 \; = \; 
\begin{pmatrix}
-J & 0 \\ 0 & J
\end{pmatrix}
\; .
$$
The algebra of real $4\times 4$ matrices so obtained is isomorphic to
the quaternionic algebra $\Hbb$. With these identifications we can
represent the algebra of octonions in terms of an algebra of $4 \times
4$ matrices whose entries are elements of the sub-algebra generated by
$(e_{0},\, e_{7})$. Denoting by $\pi(O)$ the representative of an
octonion $O$ given as (\ref{eq:7}), we have:
\begin{gather}
\pi(O) \; = \; \left[\begin{array}{cccc}
Y_{0} & -Y_{3} & -Y_{2} & -Y_{1}\\
Y_{3} & Y_{0} & Y_{1} & -Y_{2}\\
Y_{2} & -Y_{1} & Y_{0} & Y_{3}\\
Y_{1} & Y_{2} & -Y_{3} & Y_{0}
\end{array}\right].\label{eq:10}
\end{gather}
Identifying the entries $Y_{k}$, $k = 0, \; \ldots ,\;  3$, above, with complex
numbers, eq.\ (\ref{eq:10}) provides an homomorphism (not an isomorphism)
between $\Obb$ and an algebra of $4 \times 4$ complex matrices. Equation
(\ref{eq:10}) suggests, in addition, a convenient definition of
derivative operations, as we will see below.

\section{Fields and a derivative operator}

Let us now consider $Y_{k}$, $k = 0, \; \ldots ,\; 3$, as functions
$Y_{k}(t,\,x_{1},\, x_{2},\, x_{3})$ defined in $\R^{4}$ taking
values on the sub-algebra of $\Obb$ generated by $(e_0, \; e_7)$, i.e.,
taking values on the $e_0$--$e_7$ plane. Denote by $\partial_t$
partial derivative with respect to $t$ and by
 $\partial_{k}$ the
partial derivative with respect to $x_{k}$, $k = 1, \; 2, \; 3$,
and define a derivative operator $\partial$ in terms of quaternion
basis elements as equation (\ref{eq:8}),
\begin{equation}
\partial \; = \; \partial_{t}e_{0}+\partial_{1}e_{1}+\partial_{2}e_{2}+\partial_{3}e_{3} \;.
\label{eq:11}
\end{equation}
By analogy with (\ref{eq:10}), the derivative \ensuremath{\partial} can be written
in terms of $4 \times 4$ matrix as
\begin{gather}
\partial=\left[\begin{array}{cccc}
\partial_{t} & -\partial_{3} & -\partial_{2} & -\partial_{1}\\
\partial_{3} & \partial_{t} & \partial_{1} & -\partial_{2}\\
\partial_{2} & -\partial_{1} & \partial_{t} & \partial_{3}\\
\partial_{1} & \partial_{2} & -\partial_{3} & \partial_{t}
\end{array}\right].\label{eq:12}
\end{gather}
Consider now a field of octonions $\R^{4}\ni(t, \; x_1, \; x_2, \;
x_3)\mapsto O(t, \; x_1, \; x_2, \; x_3)\in \Obb$ in the representation
(\ref{eq:10}). The matrix representation (\ref{eq:12}) of the
derivative $\partial$ suggests to define the derivative $\partial \pi(O)$
of a octonionic field $O$ by taking the matrix product of
(\ref{eq:12}) with (\ref{eq:10}):

{\tiny 
\begin{gather}
\partial \pi(O)=\left[\begin{array}{cccc}
\partial_{t}Y_{0}-\partial_{1}Y_{1}-\partial_{2}Y_{2}-\partial_{3}Y_{3} & -\partial_{3}Y_{0}+\partial_{2}Y_{1}-\partial_{1}Y_{2}-\partial_{t}Y_{3} & -\partial_{2}Y_{0}-\partial_{3}Y_{1}-\partial_{t}Y_{2}+\partial_{1}Y_{3} & -\partial_{1}Y_{0}-\partial_{t}Y_{1}+\partial_{3}Y_{2}-\partial_{2}Y_{3}\\
\partial_{3}Y_{0}-\partial_{2}Y_{1}+\partial_{1}Y_{2}+\partial_{t}Y_{3} & \partial_{t}Y_{0}-\partial_{1}Y_{1}-\partial_{2}Y_{2}-\partial_{3}Y_{3} & \partial_{1}Y_{0}+\partial_{t}Y_{1}-\partial_{3}Y_{2}+\partial_{2}Y_{3} & -\partial_{2}Y_{0}-\partial_{3}Y_{1}-\partial_{t}Y_{2}+\partial_{1}Y_{3}\\
\partial_{2}Y_{0}+\partial_{3}Y_{1}+\partial_{t}Y_{2}-\partial_{1}y_{3} & -\partial_{1}Y_{0}-\partial_{t}Y_{1}+\partial_{3}Y_{2}-\partial_{2}Y_{3} & \partial_{t}Y_{0}-\partial_{1}Y_{1}-\partial_{2}Y_{2}-\partial_{3}Y_{3} & \partial_{3}Y_{0}-\partial_{2}Y_{1}+\partial_{1}Y_{2}+\partial_{t}Y_{3}\\
\partial_{1}Y_{0}+\partial_{t}Y_{1}-\partial_{3}Y_{2}+\partial_{2}Y_{3} & \partial_{2}Y_{0}+\partial_{3}Y_{1}+\partial_{t}Y_{2}-\partial_{1}Y_{3} & -\partial_{3}Y_{0}+\partial_{2}Y_{1}-\partial_{1}Y_{2}-\partial_{t}Y_{3} & \partial_{t}Y_{0}-\partial_{1}Y_{1}-\partial_{2}Y_{2}-\partial_{3}Y_{3}
\end{array}\right].\label{eq:13}
\end{gather}
}

The matrix elements in equation (\ref{eq:13}) are analogous to the
corresponding matrix elements in equation (\ref{eq:10}), that has a
compact form as equation (\ref{eq:7}). This suggests to define a
derivative operator on an octonionic field $O(t, \; x_1, \; x_2, \;
x_3)$ by imposing $\pi\big(\partial O\big)=\partial\pi(O)$. One gets,
\begin{align}
\partial O= & \left[\partial_{t}Y_{0}-\partial_{1}Y_{1}-\partial_{2}Y_{2}-\partial_{3}Y_{3}\right]e_{0}+\left[\partial_{1}Y_{0}+\partial_{t}Y_{1}+\partial_{2}Y_{3}-\partial_{3}Y_{2}\right]e_{1}+\nonumber \\
 & \left[\partial_{2}Y_{0}+\partial_{t}Y_{2}+\partial_{3}Y_{1}-\partial_{1}Y_{3}\right]e_{2}+\left[\partial_{3}Y_{0}+\partial_{t}Y_{3}+\partial_{1}Y_{2}-\partial_{2}Y_{1}\right]e_{3}\; \in \; \Obb\;, \label{eq:14}
\end{align}
or, in a more compact form, 
\begin{align}
\partial O= &
[\partial_{t}Y_{0}-\partial_{j}Y_{j}]e_{0}+\left[\partial_{j}Y_{0}+\partial_{t}Y_{j}+\left(\overrightarrow{\nabla}\times\overrightarrow{Y}\right)_{j}\right]e_{j}
\; ,
\label{eq:15}
\end{align}
where $\overrightarrow{Y}$ is the $3$-vector with components $(Y_1, \;
Y_2, \; Y_3)$.  Above we used Einstein's summation convention (the
summation over $j$ being from $1$ to $3$).  This expression will be
used to describe the fully symmetric Maxwell's equations in the next
section.

\section{Generalised Maxwell's equations generated by octonions}

Until now we considered the functions $Y_k$, $k=0, \, \ldots , \, 3$,
as functions of $t$ and of space coordinates $x_j$, $j=1, \; 2 ,
\; 3$. The $e_0$--$e_7$ plane in the octonions algebra $\Obb$ is a
sub-algebra isomorph to the algebra of the complex numbers $\C$.  Our
next step is to consider $t$ as an octonionic variable $t\mapsto e_0t
$ and to perform an analytic continuation of the functions $Y_k$, to
this complex plane, by performing a sort of Wick rotation $
e_0t\mapsto e_7 x_0$ (with $x_0\in\R$). Formally, by this procedure of
analytic continuation of the functions $Y_k\big(t, \; \vec{x}\big)$
remain functions taking values on the $e_0$--$e_7$ plane, i.e., we
may, as before, write
$$
Y_k\big( e_7 x_0, \; \vec{x}\big)
\; = \; 
e_0 Y_k^{(0)}\big( x_0, \; \vec{x}\big)
+
e_7 Y_k^{(1)}\big( x_0, \; \vec{x}\big)
\;,
$$
with $Y_k^{(0)}$ and $Y_k^{(1)}$ being real function analogous to the real and
imaginary parts of $Y_k$. 

By this procedure, the time derivatives $\partial_t$ are replaced in (\ref{eq:14})
by $-e_7 \partial_{x_0}\equiv -e_7\partial_0$. Hence, (\ref{eq:15}) becomes
\begin{eqnarray}
\partial O & = &
\left[
-e_7\partial_0\left(Y_{0}^{(0)}+e_7Y_{0}^{(1)}\right)-\partial_{j}\left(Y_{j}^{(0)}+e_7Y_{j}^{(1)}\right)
\right]e_{0}
\nonumber \\
& & 
+ 
\left[\partial_{j}\left(Y_{0}^{(0)}+e_7Y_{0}^{(1)}\right)-e_7\partial_0\left(Y_{j}^{(0)}+e_7Y_{j}^{(1)}\right)+
\left(\overrightarrow{\nabla}\times\left(\overrightarrow{Y^{(0)}}+e_7\overrightarrow{Y^{(1)}}\right)\right)_{j}\right]e_{j}
\nonumber \\
 & = &
e_0\left[ 
  \partial_0Y_{0}^{(1)}  - \partial_{j}Y_{j}^{(0)}
  + \left( \partial_{j}Y_{0}^{(0)}  + \partial_0Y_{j}^{(1)}+
  \left(\overrightarrow{\nabla}\times\overrightarrow{Y^{(0)}}\right)_j\right)e_j
\right]
\nonumber \\
& &
+ e_7 \left[ 
  -\partial_0Y_{0}^{(0)}  - \partial_{j}Y_{j}^{(1)}
  + \left(\partial_{j}Y_{0}^{(1)} - \partial_0Y_{j}^{(0)} +
  \left(\overrightarrow{\nabla}\times\overrightarrow{Y^{(1)}}\right)_j\right)e_j
\right] \;.
\label{eq:15-b}
\end{eqnarray}

Now, we change our notations by writing
\begin{eqnarray}
\overrightarrow{Y} & = & \overrightarrow{Y^{(0)}} +
e_7\overrightarrow{Y^{(1)}}
\;\; \equiv \;\;
 \overrightarrow{E}+e_{7}\overrightarrow{H} \; ,
\label{eq:not-change-a}
\\
\partial_{0}Y_{0}   & = &  \partial_{0}Y_0^{(0)} +
e_7 \partial_{0}Y_0^{(1)}
\;\; \equiv \;\;
-\rho_{m}+e_{7}\rho_{e}\; ,
\label{eq:not-change-b}
\\
\partial_{j}Y_{0} & = & 
\partial_{j}Y_{0}^{(0)} + e_7 \partial_{j}Y_{0}^{(1)}
\;\; \equiv \;\;
 \big(j_{m}\big)_j-e_{7}\big(j_{e}\big)_j\; ,
\quad \forall j=1, \; 2 , \; 3\;.
\label{eq:not-change-c}
\end{eqnarray}
With this new notation, eq.\ (\ref{eq:15-b}) becomes 

\begin{eqnarray}
\partial O & = &
e_0\left[ 
  \rho_e  - \partial_{j}E_j
     + \left(  (j_m)_j  + \big(\partial_{0} H_j\big)  +
  \left(\overrightarrow{\nabla}\times\overrightarrow{E}\right)_j\right)e_j
\right]
\nonumber \\
& &
+ e_7 \left[ 
  \rho_m  - \partial_{j}H_j
  + \left( -\big(j_e\big)_j - \big(\partial_0E_j\big) +
  \left(\overrightarrow{\nabla}\times\overrightarrow{H}\right)_j\right)e_j
\right] \;.
\label{eq:15-c}
\end{eqnarray}

By imposing $\partial O=0$, the different components of equation
(\ref{eq:15-c}) provide eight equations corresponding to the
coefficients of $e_{0}$, $e_{7}$, $e_{j}$ and $e_{7}e_{j}$, $j=1,\; 2
, \; 3$. These equations can be more properly written in vector form:
\begin{eqnarray}
\overrightarrow{\nabla}\cdot\overrightarrow{E} & = & \rho_{e}\; , \label{eq:19a}\\
\overrightarrow{\nabla}\cdot\overrightarrow{H} & = & \rho_{m}\; , \label{eq:19b}\\
\overrightarrow{\nabla}\times\overrightarrow{E} & = &
-\frac{\partial\overrightarrow{H}}{\partial t}-\overrightarrow{j_{m}}\; , \label{eq:19c}\\
\overrightarrow{\nabla}\times\overrightarrow{H} & = & \frac{\partial\overrightarrow{E}}{\partial t}+\overrightarrow{j_{e}}\; .\label{eq:19d}
\end{eqnarray}

As we now recognise, equations (\ref{eq:19a})--(\ref{eq:19d}) are the
fully symmetric Maxwell's equations and allow for the possibility of
magnetic charges and currents, analogous to electric charges and
currents. It is a well known fact that equations
(\ref{eq:19a})--(\ref{eq:19d}) are unchanged under the so-called
duality transformations, defined, for $\theta \in [0, \; 2\pi)$, by
$$
\begin{pmatrix}
\rho_e \\ \rho_m
\end{pmatrix} 
\; \mapsto \; R(\theta) 
\begin{pmatrix}
\rho_e \\ \rho_m
\end{pmatrix} 
\; , \quad
\begin{pmatrix}
\overrightarrow{j_{e}} \\ \overrightarrow{j_{m}}
\end{pmatrix} 
\; \mapsto \; R(\theta) 
\begin{pmatrix}
\overrightarrow{j_{e}} \\ \overrightarrow{j_{m}}
\end{pmatrix} 
\quad \mbox{ and }\quad
\begin{pmatrix}
\overrightarrow{E} \\ \overrightarrow{H}
\end{pmatrix} 
\; \mapsto \; R(\theta) 
\begin{pmatrix}
\overrightarrow{E} \\ \overrightarrow{H} 
\end{pmatrix}
$$
where
$
R(\theta)  :=  
\begin{pmatrix}
\cos\theta & -\sin\theta
\\
\sin\theta & \cos\theta
\end{pmatrix}
$. 
We should also recall that, using arguments from quantum mechanics, Dirac
\cite{key-16} argued that the existence of an isolated magnetic charge
would imply the quantisation of electric charge.

From (\ref{eq:19a})--(\ref{eq:19d}) we easily obtain the so-called
continuity equations for the electric and magnetic charge densities
and currents: $\partial_0 \rho_e +
\overrightarrow{\nabla}\cdot\overrightarrow{j_e} = 0$ and $\partial_0
\rho_m + \overrightarrow{\nabla}\cdot\overrightarrow{j_m} = 0$.  By
using (\ref{eq:not-change-b}) and (\ref{eq:not-change-c}), these
continuity equations imply, as one easily sees, $ \Box Y_o^{(0)} = 0 $
and $ \Box Y_o^{(1)} = 0 $, or simply $\Box Y_0=0$, where $\Box$ is the
D'Alembertian operator.

\section{Conclusion}

We have considered a representation of octonions as a four dimensional
algebra over complex numbers. After a suitable definition of a
derivative operation over fields of octonions in $3+1$ dimensions we
derived the fully symmetric Maxwell's equations (i.e., with electric and
magnetic charges and currents) as a single equation.

\vspace{0.5cm}

\noindent{\bf Acknowledgment.} One of the authors (P.) wishes to thank
FAPESP for financial support.

\end{document}